Construction and Evaluation of Hierarchical Parcellation of the Brain using fMRI with Prewhitening

Pantea Moghimi[*1], Kelvin O. Lim[2], Theoden I. Netoff[1]

[1]Department of Biomedical Engineering, University of Minnesota, Minneapolis, Minnesota, USA

[2]Department of Psychiatry, University of Minnesota, Minneapolis, Minnesota, USA

[*]Corresponding author: tnetoff@umn.edu

# Abstract

Brain atlases are a ubiquitous tool used for analyzing and interpreting brain imaging datasets. Traditionally, brain atlases divided the brain into regions separated by anatomical landmarks. In the last decade, several attempts have been made to parcellate the brain into regions with distinct functional activity using fMRI. To construct a brain atlas using fMRI, data driven algorithms are used to group voxels with similar functional activity together to form regions. Hierarchical clustering is one parcellation method that has been used for functional parcellation of the brain, resulting in parcellations that align well with cytoarchitectonic divisions of the brain. However, few rigorous data driven evaluations of the method have been performed. Moreover, the effect of removing autocorrelation trends from fMRI time series (prewhitening) on the structure of the resultant atlas has not been previously explored. In this paper, we use hierarchical clustering to produce functional parcellations of the brain using hierarchical clustering. We use both prewhitened and raw fMRI time series to construct the atlas. The resultant atlases were then evaluated for their homogeneity, separation between regions, reproducibility across subjects, and reproducibility across scans.

# Introduction

Brain atlases are used in analysis of functional magnetic resonance imaging (fMRI) datasets. They group voxels into contiguous regions whose functional activity is to be averaged to represent the dynamics of brain regions over time. Averaging voxels increases signal to noise ratio as well as reducing dimensionality of the dataset from thousand of voxels to dozens of regions. Time series from the regions defined by the atlases may be used for biomarkers of diseases through analysis of the time series and network analysis of the interactions between the areas.

Historically, anatomical atlases have been used for analysis of fMRI datasets. Anatomical atlases divide the brain into its major gyri using anatomical landmarks identified manually. Several anatomical atlases such the Automated Anatomical Labeling atlas (AAL) [1], Desikan-Killiany atlas [2], and Destrieux atlas [3,4] are commonly used.

Anatomical divisions of the brain may not reflect its functional organization and therefore may confound two neighboring areas with different functions and dynamics into a single area. Network analysis of functional activity based on anatomical atlases may not be as accurate or as sensitive as from functional atlases. In recent years several attempts have been made to produce parcellations of the brain that are based on functional activity of voxels, rather than their spatial location (e.g. [5–11]). These studies use data driven unsupervised methods, known as clustering, to group voxels with similar functional activity to form regions. Functional activity is typically collected in resting state, when subjects are asked to relax in the scanner and are not required to perform any tasks, although one study has used task evoked activity for parcellation [10]. The resultant parcellation, known as a functional brain atlas, can be used in place of anatomical atlases.

Functional parcellation methods previously used have several limitations, such as contiguity, size bias, and spatial bias. Some methods do not produce contiguous regions (e.g. [6,7,9] resulting in brain "regions" that are spatially scattered across the brain, complicating interpretation of any results obtained from the atlas. Some methods, such as the K-means or spectral clustering algorithms, are biased towards regions of equal size (e.g. [8,10,11]). And some methods only parcelate the cortical surface neglecting subcortical structures [5].

Grouping of voxels into regions is based on their pairwise similarity, often quantified as the pairwise cross correlation between the BOLD signal of the voxels. The presence of strong autocorrelations within each time series can cause spuriously high correlation values [12]. Removal of autocorrelation, a process known as prewhitening, may improve accuracy of parcellation based on the measured interactions by removing the bias caused by signals that have similar autocorrelations.

In this paper we develop and evaluate a functional parcellation and compare it to anatomical parcellation. To explore the role of prewhitening, we constructed two functional atlases based on the cross correlation between voxels using the raw time series, the other using pre-whitened time series. The correlation data is then used to create a functional parcellation of the brain using a spatially constrained hierarchical clustering algorithm because hierarchical clustering is not biased towards regions of equal size [13]. By adding a spatial constraint to the hierarchical clustering algorithm, this method can produce contiguous regions. Previous work has shown this hierarchical clustering results in parcellations that are stable across different scans of the same subjects and in agreement with cytoarchitectonic maps of the brain [14].

Functional atlases were generated from a set of 88 healthy subjects with 6 minute resting state fMRI scans. We evaluated the resultant atlases in terms of homogeneity of the regions, separation between regions, and reproducibility of results on the individual subjects and report average scores. We explored effect of scan duration on parcellation results. We also compared functional atlases constructed using combined datasets from our group of subjects to atlases constructed from individual scans.

# Methods

## Participants

A group of 88 (27 female, age: M = 33.4, SD = 11.9) subjects with no neurological disorders participated in this study. All participants gave informed consent and were compensated for their participation. All procedures were done in accordance with a University of Minnesota IRB approved protocol.

## Image Acquisition

Each subject underwent a six minute resting state fMRI image acquisition, during which subjects were instructed to stay still and awake and keep their eyes closed for about 6 minutes. Images were acquired using a Siemens Trio 3T scanner (Erlangen, Germany) with the following sequence parameters: gradient-echo echo-planar imaging (EPI) 180 volumes, repetition time (TR) 2 seconds, echo time (TE) 30ms, flip angle 90º, 34 contiguous AC-PC aligned axial slices, voxel size 3.4 x 3.4 x 4.0 mm, matrix 64 x 64 x 34 totalling 139,264 voxels.

In addition to functional activity, a T1-weighted anatomical image was acquired using a magnetization prepared rapid gradient-echo sequence. A field map was also acquired and used to correct for geometric distortions introduced by field inhomogeneities: TR = 300ms, TE = 1.91 ms/4.37 ms, flip angle = 55º, voxels size = 3.4 x 3.4 x 4.0 mm [15,16].

The raw fMRI data was preprocessed using FEAT and MELODIC from the FSL software package as follows. The first three volumes were removed from each subject scan to account for magnetization stabilization. This resulted in a 5.9 minute time series per voxel (177 time points). Each scan was motion corrected, B0 field map unwarped, and corrected for slice scan time. Non-brain portions of the images were removed and a spatial smoothing kernel was applied to the dataset (6mm full-width half-maximum). The images were then grand mean and intensity normalized and temporally filtered between 0.01 and 0.08Hz. All images were then registered to the MNI152 space. Using probabilistic independent component analysis (PICA), noise introduced by head motion, respiration, cardiac pulsation, and scanner artifacts was removed. Spatial and temporal characteristics of noise components are described in MELODIC manual (https://fsl.fmrib.ox.ac.uk/fslcourse/lectures/melodic.pdf). The dataset was then resampled to 3 x 3 x 3mm, resulting in 47640 voxels.

## Prewhitening

Prewhitening refers to removal of autocorrelation from a given time series so that similar to white noise, the resultant time series are decorrelated. Presence of autocorrelation in BOLD time series can lead to spurious high cross correlation values between different voxels [12]. We prewhitened the time series from voxel $i$, $x_i(t)$ by calculating its Fourier transform $X_i(f)$ and dividing it by its power spectrum $|X_i(f)|$, to result in a flat power spectrum, similar to white noise. The resultant signal $X_i^W(f)$ was then inverse Fourier transformed into the time domain $x_i^W(t)$ (Equation 1).

$$x_i(t) \leftrightarrow X_i(f)$$

$$X_i^W(f) = X_i(f)/|X_i(f)| \qquad \text{Equation (1)}$$

$$x_i^W(t) \leftrightarrow X_i^W(f)$$

## Parcellation

To parcellate the brain using fMRI data, voxels with similar time series are grouped together to form regions. This is typically done using data-driven clustering algorithms, where each cluster constitutes one region (e.g. [8,10,11]).

We chose the agglomerative hierarchical clustering algorithm with Ward's minimum variance as linkage criterion [13,17]. Hierarchical clustering algorithm is not biased towards clusters of equal size like the K-means or spectral clustering algorithms [13] and results in more reproducible parcellations [10]. The agglomerative hierarchical clustering algorithm starts with each datapoint (voxel) as a single cluster. It then merges the cluster pair that minimizes Ward's criterion to form a new cluster. Ward's criterion calculates total within-cluster variance resulting from merging each pair of clusters. The merging process is iterated until all clusters are merged to form a single cluster containing all data points. Information about membership of each datapoint to each cluster at each stage of merging is stored in a structure called a dendrogram. Different parcellation scales, i.e. number of regions the brain is parcellated into, are constructed by cutting the dendrogram at the stage that contain the desired number of clusters. The resultant parcellations are then used as functional brain atlases. To obtain atlases with contiguous regions, we applied a spatial constraint so that only spatially adjacent clusters can be merged.

A clustering algorithm requires a distance measure between voxel pairs. We used the correlation distance for parcellation. Correlation distance between voxels $i$ and $j$, $d_{i,j}$, is equal to $d_{i,j} = 1 - r_{i,j}$ where $r_{i,j}$ is the zero-lag cross correlation between the two voxels.

We constructed two types of functional atlas: i) atlas constructed using the original time series, referred to as original functional atlas in this manuscript; ii) atlas constructed using prewhitened time series, referred to as white functional atlas in this manuscript.

We constructed functional atlases at two different levels: i) Group level, where time series from the entire group of subjects were combined to construct a group dataset, which was then used for construction of the functional atlas. To combine individual datasets, for each voxel, time series from all subjects were concatenated to construct a single time series. ii) Individual level, where the functional atlas was constructed for each individual dataset. In this manuscript, unless stated otherwise, functional atlas refers to a group level functional atlas.

## Evaluation

To evaluate the resultant functional atlases we employed three approaches: i) calculating homogeneity of the regions, which measures how similar voxels within a single region are to each other; ii) calculating separation between regions, which measures how dissimilar voxels in different regions are with respect to voxels within regions; iii) calculating reproducibility of parcellation results, which quantifies how reproducible the results are if a different group of subjects, to different datasets from the same subject were used for parcellation. Each approach is explained here in detail. Parcellations were generated based on group level data, but the performance of the parcellation was measured for each subject. Average performance scores of the functional atlases vs. anatomical and random atlases are compared.

## Homogeneity

Several measures of homogeneity have been used to evaluate parcellation methods. We used the following 5 measures to quantify homogeneity of the resultant atlases:

1. Average pairwise correlation coefficient between voxels within each region [8,11], referred to as *rt* in this manuscript.
2. Average pairwise correlation coefficient between functional connectivity maps between voxels within each region [8], referred to as *rs* in this manuscript.
3. Percentage of variance explained by the first principal component [18] of time series of voxels within each region [19], referred to as *pcat* in this manuscript.
4. Percentage of variance explained by the first principal component of functional connectivity maps of voxels within each region, referred to as *pcas* in this manuscript.
5. Kendall's coefficient of concordance [20] between voxels within each region [21], referred to as *KCC* in this manuscript.

Due to the spatial autocorrelation present in fMRI datasets, a contiguous random grouping of voxels is bound to result in regions with a certain degree of homogeneity. Therefore, to test our null hypothesis we compared distribution of homogeneity of the functional atlases to homogeneity of randomly constructed atlases with contiguous regions and similar size distributions. Homogeneity depends on size of regions. Smaller regions contain fewer voxels which results in less diversity among the voxels. The extreme case is a region that consists of a single voxel, which is perfectly homogeneous. Therefore, the random atlases must match the functional atlases in distribution of region sizes. We constructed random atlases that consisted of spatially contiguous regions with similar region size distribution to functional atlases, but assignment of voxels to regions was performed randomly. To construct a random atlas with M regions, we randomly picked M initial voxels as seeds, with each seed constituting a single region. Pairwise Euclidean distance between each of the voxels and the seed voxels was calculated and each voxel was assigned to the region with the closest seed. Since this algorithm does not guarantee the distribution of region sizes to match that of our atlases, 1000 random atlases were generated and the mismatch of their size distribution to the functional atlases was calculated. Then the 10 random atlases with lowest mismatch were used as the final random atlases. Average mismatch for the 10 chosen atlases was less than 5%.

## Cluster Separation

To quantify separation between regions, we used the Silhouette coefficient [22] which has been used in several studies to evaluate parcellation algorithms [7,8,23–25]. The silhouette coefficient measures how similar each voxel is to voxels within its region compared to voxels in other regions. To calculate Silhouette coefficient for voxel $i$, first, average correlation distance, $a_i$, between voxel $i$ and all other voxels assigned to the same region. Then, the lowest average correlation distance between that voxel and all other regions, $b_i$ is calculated, where average distance between the voxel and each region is average distance between that voxel and all the voxel belonging to that region. The silhouette coefficient for voxel $i$ is then calculated as $S_i = (b_i - a_i)/max(b_i, a_i)$. Silhouette coefficient takes up values between 1 and -1, where a value of 1 indicates that the region the voxel belongs to is well separated from other regions.

Similar to homogeneity values, Silhouette coefficient values from the functional atlases were compared to values from randomly constructed atlases.

Reproducibility

To calculate how reproducible the results of functional parcellation are across different groups, we divided our subjects into two groups, and constructed separate functional atlases from each group's raw dataset. We then compared the agreement between the two atlases. This comparison was done at four different parcellation scales, 90, 500, 1000, and 4000 regions. We quantified the agreement using the Adjusted Rand Index (ARI), a measure of comparing different groupings of the same dataset [26,27]. Rand index (RI) is a normalized measure that calculates agreement between two parcellations (Equation 2). ARI is a corrected form of RI that subtracts expected RI values that are to be observed due to chance. ARI can take up values between 1 (total agreement between the two parcellations) and -1 (total disagreement between the two parcellations).

$a$: Total number of voxels pairs that are assigned to the same region in both parcellations

$b$: Total number of voxels pairs that are assigned to different regions in both parcellations

$c$: Total number of voxels pairs that are assigned to the same region in parcellation 1 and to different regions in parcellation 2

$d$: Total number of voxels pairs that are assigned to different regions in parcellation 1 and to the same region in parcellation 2

$RI = \frac{a+b}{a+b+c+d}$  *Equation 2*

## Scan duration

To evaluate effect of scan duration on homogeneity and reproducibility of the functional atlas, we constructed functional atlases using a range of scan durations. A subset of our subjects (N=24), were scanned for a second and third time in six and nine months after the first scan. Each scan session lasted 5.9 minutes. Raw time series for each voxel were concatenated for each subject to construct a 17.7 minutes long time series per voxel.

To examine reproducibility of parcellation results as a function of scan duration, we divided the long time series into two halves. 360s and 600s from each half were taken and used to construct functional atlases with 90, 500, and 2000 regions.

To quantify the effect of scan duration on regional homogeneity, the 17.7 minute long time series were truncated at different time points, and truncated time series were used separately to construct functional brain atlases at two different scales, 90 and 500 regions. Time series were truncated after the first 360s (6 minutes), 600s (10 minutes), 840s (14 minutes) and 1062s (17.7 minutes). Homogeneity of the regions of each functional atlas was then calculated.

## Level of Analysis

We constructed individual level functional atlases from raw time series of 7 of our subjects by parcellating each dataset separately. We quantified degree of agreement between each individual level atlas and our group level atlas using ARI at three different parcellation scales, 90, 500 and 2000 regions. We also calculated ARI between each pair of individual level atlases at those parcellation scales.

# Results

We constructed two functional atlases from fMRI data combined across all the subjects into one dataset. Given the group level cross correlation data between voxels, we used the agglomerative hierarchical clustering with the linkage method for merging criterion. The first functional atlas, referred to as the original functional atlas, was constructed using raw time series. The second functional atlas, referred to as the white functional atlas, was constructed using pre-whitened time series. Pre-whitening removes autocorrelation from each time series, resulting in elimination of spurious cross correlation between voxels [12]. We observed that average pairwise cross correlation between all voxels was reduced after pre-whitening (Figure 1).

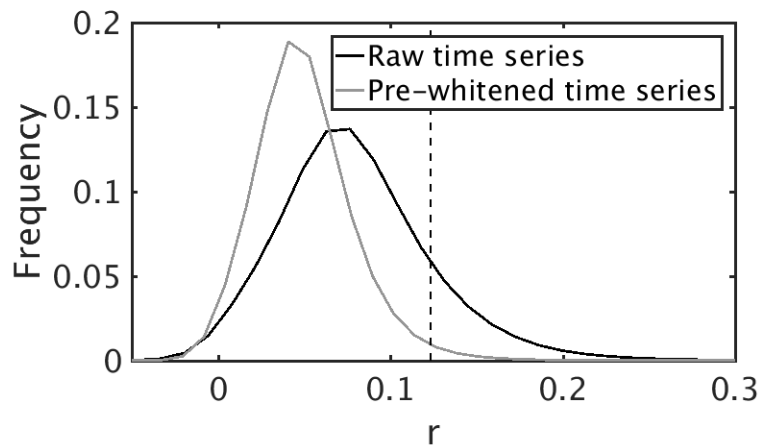

**Figure 1. Distribution of Pairwise Cross Correlations**

Distribution of pairwise cross correlation between voxels for raw and pre-whitened time series. Vertical dash line marks significance threshold at $p<0.05$. Correlation values higher than the threshold are considered significant. Prewhitening reduces number of pairwise correlations considered "significant" by removing spurious correlations.

To compare properties of the functional atlases against anatomical atlas, we constructed each functional atlas with 90 regions to compare it to a commonly used anatomical atlas, known as the AAL, which also consists of 90 regions. Schematics of the resultant functional atlases along with the AAL anatomical atlas [1] are shown in Figure 2A, 2B, and 2C.

Distribution of region sizes for the three atlases are shown in Figure 2D. As can be seen in Figure 2D left, the two functional atlases have similar size distributions compared to the AAL atlas. However the AAL atlas has regions with smaller sizes than the functional atlases. We also constructed ten random parcellations that matched each functional atlas in size distribution. Properties of random atlases were compared to that of the functional atlases. These random parcellations match the region size distribution of the functional atlases very closely.

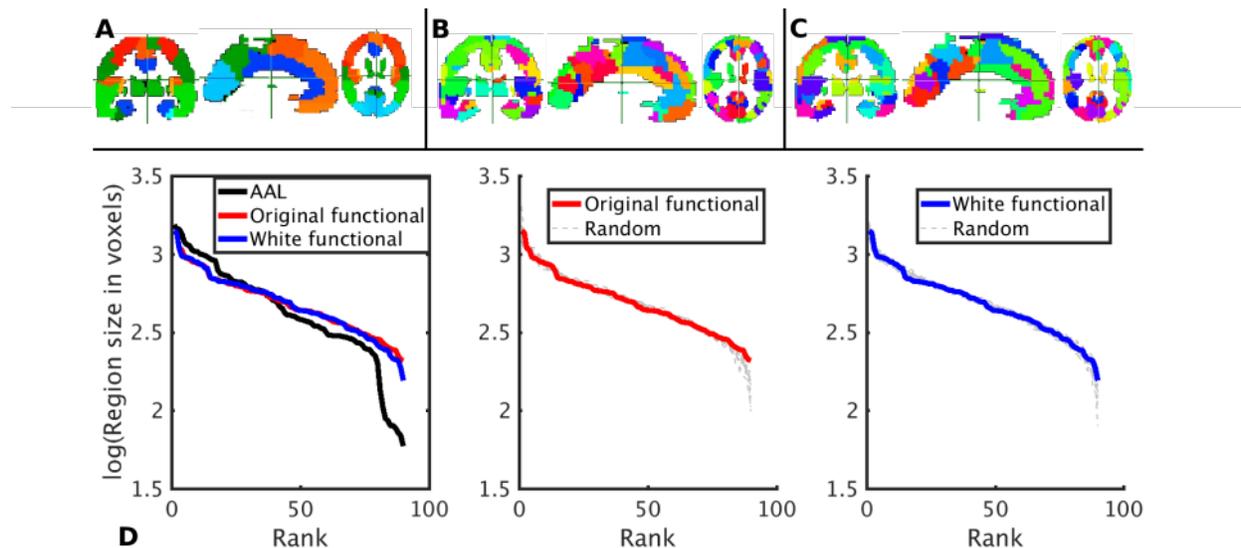

**Figure 2. Parcellation Properties**

(A)  Parcellation using the AAL atlas, coronal (left), sagittal (middle), and horizontal (right) views

(B)   Parcellation using the original functional atlas, constructed using raw time series.

(C)   Parcellation using white functional atlas, constructed using prewhitened time series

(D)   Size distributions. Regions were rank ordered based on their size. X-axis: Rank of each region, Y-axis: Number of voxels within the region in log scale. Left: Distribution of size of regions for the AAL, original functional, and white functional atlases. Middle: Distribution of size of regions for the original functional atlas constructed using raw time series, as well as ten size matched random parcellations. Right: Distribution of size of regions for the white functional atlas constructed using prewhitened time series, as well as ten size matched random parcellations.

We also tested other linkage criteria. However, other linkage criteria resulted in atlases with several regions of single voxels as well as regions that were extremely large, encompassing entire lobes. The Ward linkage criterion produced region sizes the most comparable to the AAL (Figure 3).

Next, we tested whether the functional atlases was able to generalize and outperform the anatomical and random atlases using other measures of homogeneity that were not explicitly optimized for when constructing the functional atlas. Therefore, we calculated homogeneity using several other measures: *rs*, *pcat*, *pcas*, and *kcc*, the results of which are shown in Figure 4. We observed that the original and white functional atlases resulted in significantly higher *rs* and *pcat* homogeneity than AAL and random atlases but not for *pcas* or *kcc*. Homogeneity measures *rt* and *pcat* measure similarity of signals within the regions, while *rs* and *pcas* measure the similarity of the interactions with the entire brain. It appears that the homogeneity measures that measure within region voxel interactions is improved more than the homogeneity measures that are inclusive of all brain interactions.

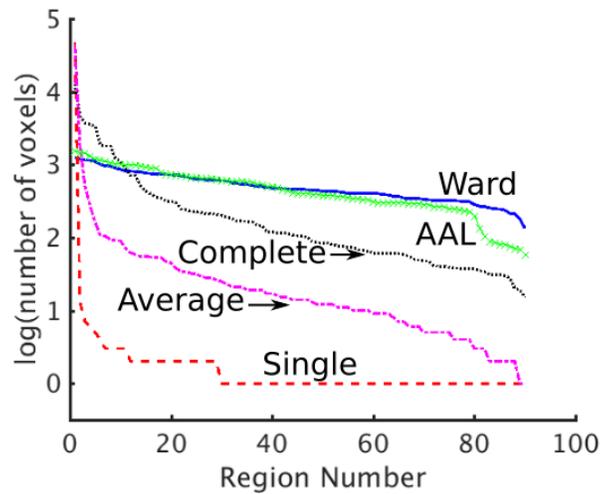

**Figure 3. Distribution of Region Sizes for Different Linkage Methods**

Distribution of size of regions for original functional atlases constructed using different linkage criteria at a scale of 90 regions as well as distribution of region sizes for the AAL atlas. Size of the regions are sorted from highest to lowest. X-axis: rank of the region, Y-axis: number of voxels in the region in log scale. Raw time series were used for construction of these atlases.

We also looked at separation between regions, which quantifies how similar voxels are between regions compared to voxels within the same region. Separation between regions was measured using the Silhouette coefficient [22]. Distribution of Silhouette values for both functional atlases, AAL atlas and random atlases are shown in Figure 4F. No significant difference between functional atlas and AAL was observed for raw (Kolmogorov-Smirnov test, p = 0.78) or prewhitened data (Kolmogorov-Smirnov test, p = 0.08) or when compared to the random atlases. Difference between Silhouette values of the white functional atlas was trending to significance, while no differences were observed in the original functional atlas.

Subsequently, we examined the degree of agreement between each pair of atlases using the Adjusted Rand Index (ARI). The results are shown in Table 1. The functional atlases calculated on raw and pre-whitened data are in 89% agreement between each other. However, the AAL atlas and each of the functional atlases have much lower agreement (about 25%). Collectively, these results show that prewhitening of the time series does not have a large effect on structure of the resultant functional atlas. But the functional atlases and the AAL atlas have considerable differences, even though they are more or less similar in number and size of their regions.

We also examined reproducibility of our results across subjects by dividing our subject set into two groups of equal size and constructing a functional atlas using the raw time series from each group separately. We then calculated degree of agreement between the two resultant atlases at several parcellation scales, using ARI. The results are shown in Figure 5. The degree of agreement between the two atlases is maximum at 90 regions, decreases as we increase the number of regions to 1000, and then increases as we further parcellate to 4000 regions. However, the difference between the maximum and minimum ARI is only 3% and the differences are probably negligible. Our conclusion from these findings is that reproducibility is not dependent on parcellation scale and that functional parcellation only moderately generalizes to new subjects.

| Pair of Atlases | ARI |
|---|---|
| Original- vs. white functional atlas | 0.89 |
| AAL vs. original functional atlas | 0.25 |
| AAL. vs. white functional atlas | 0.26 |

**Table 1. Degree of Agreement between each pair of Atlases.** Degree of agreement between each pair of atlases is quantified using the adjusted rand index (ARI).

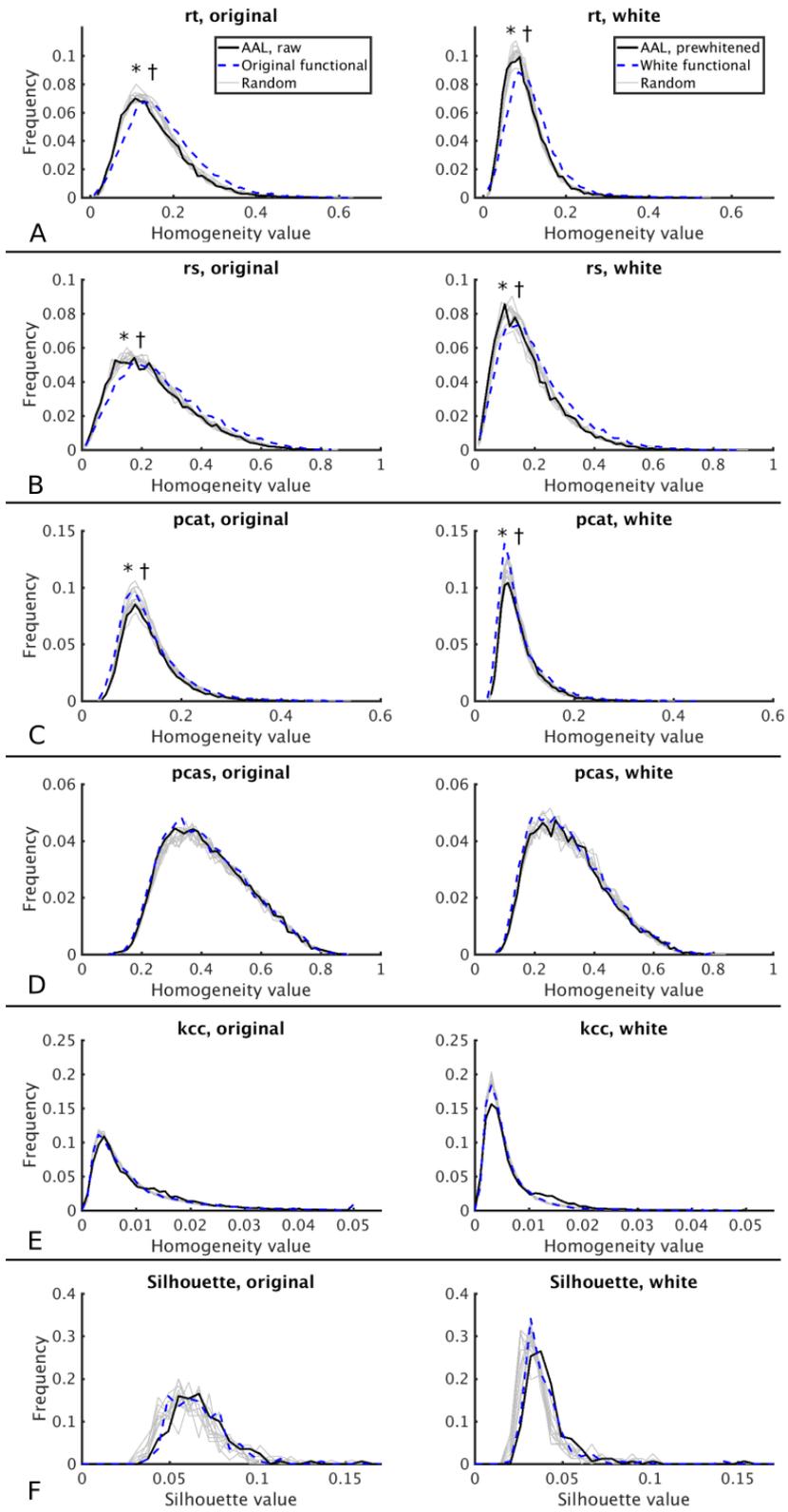

**Figure 4. Parcellation Evaluation**

(A) Distribution of homogeneity values for each atlas, measured as average pairwise correlation coefficients between voxels in each region (*rt*) across all regions and all subjects. Left: Distribution of homogeneity values for the AAL atlas applied raw time series, the original functional atlas constructed using raw time series applied to raw time series, and and random atlases size matched to the functional atlas applied to raw time series. The * indicates statistically significant difference between the original functional atlas and the AAL atlas. The † indicates statistically significant difference between the original functional atlas and each of the random atlases. Right: Distribution of homogeneity values for the AAL atlas applied prewhitened time series, the white functional atlas constructed using prewhitened time series applied to prewhitened time series, and and random atlases size matched to the functional atlas applied to prewhitened time series. The * indicates statistically significant difference between the original functional atlas and the AAL atlas. The † indicates statistically significant difference between the original functional atlas and each of the random atlases.

(B) Distribution of homogeneity values for each atlas, measured as average pairwise correlation coefficients between functional connectivity maps of voxels in each region (*rs*) across all regions and all subjects.

(C) Distribution of homogeneity values for each atlas, measured as percentage of variance explained by the first principal component of time series of all the voxels in each region (*pcat*) across all regions and all subjects.

(D) Distribution of homogeneity values for each atlas, measured as percentage of variance explained by the first principal component of functional connectivity maps of all the voxels in each region (*pcat*) across all regions and all subjects.

(E) Distribution of homogeneity values for each atlas, measured as the Kendall's Coefficient of Concordance (*kcc*) across all regions and all subjects.

(F) Distribution of Silhouette Coefficients for each atlas across subjects. For each subject, silhouette values across all regions were averaged.

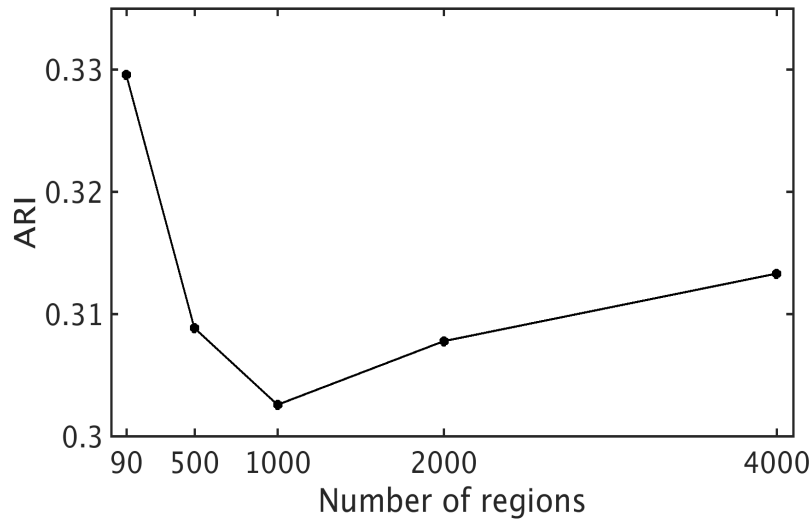

**Figure 5. Reproducibility of the Functional Parcellation across Subjects**

Agreement between two group level functional atlases based on raw data at different parcellation scales. X-axis shows scale of parcellation, i.e. number of regions. Y-axis is level of agreement between the two atlases quantified using ARI.

We then examined how reproducibility changes as a function of scan duration. For 24 subjects we had a second and third set of fMRI scans, taken 6 and 9 months after the first scan. Each scan lasted 5.9 minutes. We constructed longer time series for each voxel by concatenating the time series, resulting in 1062s (17.7 minutes) of data. We then divided the long time series into two equal halves. Then, several functional atlases were constructed by taking epochs of different durations from each half. The degree of agreement between the functional atlases constructed using the different halves was quantified using ARI. The results are shown in Figure 6A. This increase was largest for 90 regions (7% increase) compared to 500 and 2000 regions (~2% increase). We find that increasing scan duration slightly improves the reproducibility of results across datasets.

In addition, we examined *rt* homogeneity of parcellation as a function of scan duration used to construct the parcellation (Figure 6B). Parcellation at a finer scale (500 regions) results in more homogeneous regions than the coarser scale (90 regions). But, the scan duration, of the length scales tested here, does not significantly affect homogeneity of the regions.

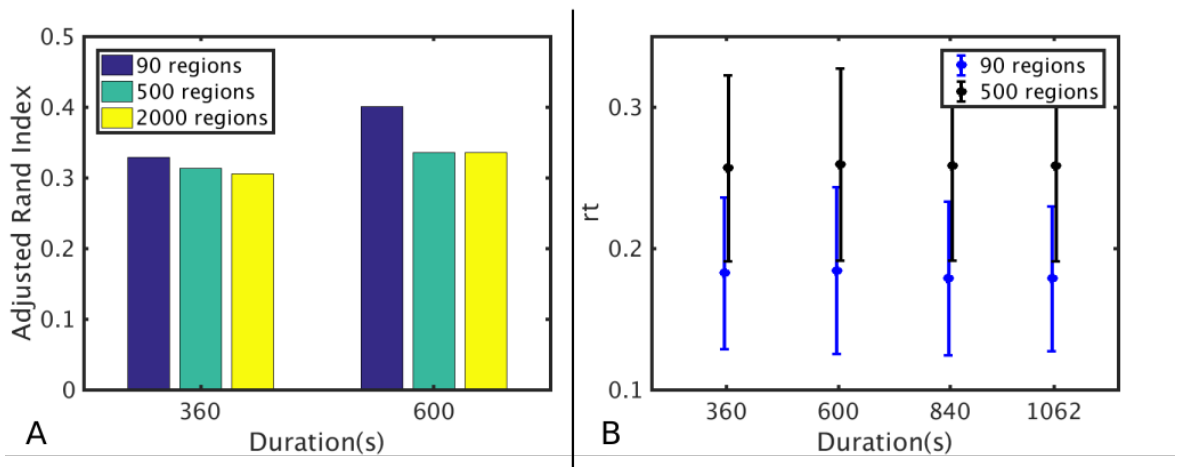

**Figure 6. Effect of Scan Duration on Homogeneity and Reproducibility**

A) Agreement between original functional atlases constructed using different datasets of equal duration, at different durations and parcellation scales.

B) Average and standard deviation of homogeneity, quantified using *rt*, across across all regions vs. scan duration used for construction of the atlas.

Lastly, we tested reproducibility of parcellations generated on individuals and subsets of subjects. We constructed individual level atlases for 7 subjects selected at random. Only seven were used because of the very long computational time it takes to construct a parcellation. We then calculated ARI between all pairs of the individual atlases (21 comparisons) as well as ARI between the individuals to the group level atlas. Results are shown in Figure 7. The degree of agreement between the individual and group level atlases are higher than between the individual level atlases. As the number of regions is increased the ARI values increases.

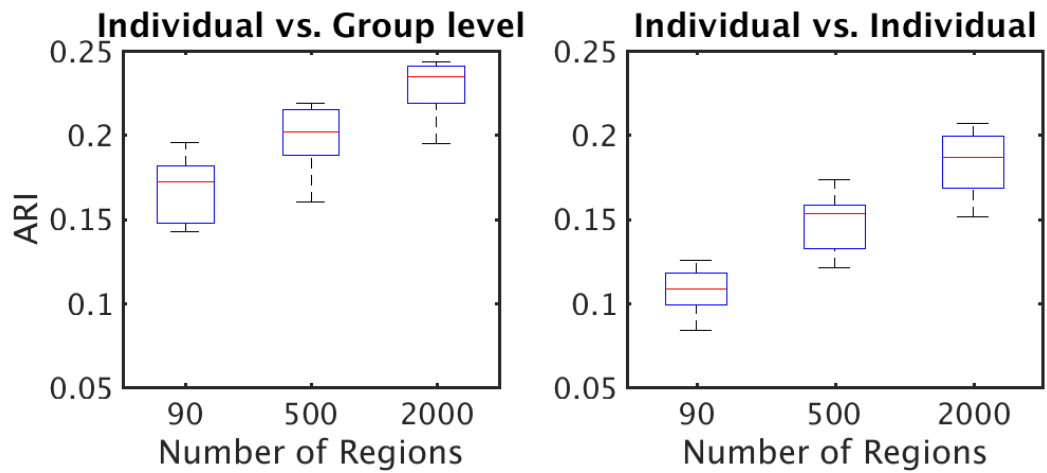

**Figure 7. Individual vs. Group Level Atlases**

Top: Distribution of degree of agreement between the seven individual level atlases and the group level atlas at different parcellation scales.

Bottom: Distributions of degree of agreement between each pair of the seven individual atlases (21 pairs total) at different parcellation scales.

# Discussion

We constructed and evaluated group level functional atlases using both raw and pre-whitened time series using the hierarchical clustering algorithm. We compared the resultant atlases with anatomical atlases. We also evaluated the resultant atlases by quantifying their homogeneity, separation between between regions, and reproducibility for group level performance. Lastly, we characterized the effect of scan duration on homogeneity and reproducibility, and compared group and individual level atlases.

Our functional atlases were similar to the anatomical atlas in terms of distribution of size of the regions, even though they were only selected to match in number of regions. Compared to the anatomical atlas and random parcellations, both functional atlases resulted in significantly more homogeneous regions when applied to individual data sets (Figure 4). This demonstrates the effectiveness of the functional atlas in grouping voxels with similar functional activity together into more homogenous regions. This significant improvement in homogeneity was observed in three out of the five homogeneity measures we used: *rt*, *rs*, and *pcat*. We expected the *rt* measure of homogeneity to have a significant improvement in functional atlases over the AAL and random atlases because they were constructed by grouping voxels to maximize cross correlation values within the region. The functional parcellation also generalized to an improvement in homogeneity quantified by *rs* homogeneity, measuring correlation between voxels within the regions to all other voxels, and *pcat*, which measured reduced complexity within the region. However, given the sample size, *pcas* and *kcc* measure of homogeneity did not reach significance. Collectively, these results demonstrate that different measures of homogeneity have different statistical powers in detecting an improvement.

Separation between the regions, quantified using the Silhouette coefficient, showed no difference between the functional atlases and the anatomical atlas or random parcellations. Although we observed that the difference between distribution of Silhouette values between the white functional atlas and the AAL atlas was close to significance (p=0.08), suggesting that using prewhitening time series did result in a functional atlas with improved separation between regions when compared to the AAL atlas.

We observed that pre-whitening results in regions with lower homogeneity and Silhouette coefficient values. This is expected, since pre-whitening removes spuriously high correlations. The white functional atlas was in high agreement with the original functional atlas (89%). This shows that even though pre-whitening reduces pairwise correlations between voxels, overall it does not affect the spatial pattern of pairwise cross correlations. However, the degree of agreement between the functional atlases and the anatomical atlas was drastically lower (~25%), demonstrating that regions delineated by functional activity do not align very well with anatomically marked regions. A similar observation was made by another study that constructed a parcellation of the brain using the pattern of white matter connections imaged by diffusion MRI. They compared the degree of agreement between their atlas and a multi-modal atlas that was constructed by Glasser and colleagues using a combination of fMRI, myelin maps, and cortical thickness [28] and observed only 28% agreement between the two atlases [29].

When the subject pool was divided into two groups and original functional atlases were generated from each group and compared, the degree of similarity between the two groups, as measured with ARI, was only slightly above 30% (Figure 5). The 30% reproducibility across group level atlases were reported in another study with 128 subjects [10]. However, using 200 subjects Arslan and colleagues reported 45% reproducibility across groups [30]. Poor reproducibility indicates that larger groups (greater than 200 subjects) are required to construct robust group level functional atlases. However, a recent study constructing brain atlases combining multiple imaging modalities reported 90% reproducibility across groups of 210 subjects.[28]

[28]eproducibility did not change drastically as a function of parcellation scale. The lack of a dependency between parcellation scale and reproducibility has also been reported in several other studies [9,10,30]. Reproducibility of the results reflects degree of agreement between functional connectivity pattern of two group datasets, which is captured by the dendrogram. As a result, the scale of parcellation, which determined by where the dendrogram is cut, does not seem to impact the result.

Reproducibility of the functional atlases across datasets was calculated as degree of agreement between functional atlases constructed from different sets from the same pool of subjects, and was comparable to reproducibility across groups (~30%). Reproducibility across datasets improved as scan duration increased (Figure 6A). This finding supports another study that demonstrated reproducibility across datasets increases as scan duration increases and approaches maximum value at about 30 minutes of data [31]. In fact, Arslan and colleagues have reported 45% reproducibility using 60 minutes resting fMRI [30].

Longer scans used for construction of the functional atlas resulted in higher reproducibility of the spatial segmentation across scans within same subjects (Figure 6A), but it did not affect homogeneity of the regions (Figure 6B). It is possible that improvements to homogeneity will occur with scans of longer durations. We did observe that increasing parcellation scale, increases region homogeneity, which is to be expected, since higher parcellation scales result in smaller regions. Smaller regions have higher homogeneity values than larger regions, since they contain a smaller number of voxels. Similar results have been reported [30].

We observed that individual level atlases have a higher degree of agreement with the group level functional atlas than individual level atlases have with each other. So the group level atlas can be thought of an "average" functional atlas that represents commonalities between the group of subjects (Figure 7). Moreover, reproducibility of individual level atlases (~15%, Figure 7) is lower than that of group level atlases (~30%, Figure 5), an observation also replicated in multimodal atlases [28]. We hypothesize that group level atlases reflect structure of the average brain and therefore are less affected by interindividual variabilities.

# Conclusion

We constructed and evaluated a functional parcellation of the brain using spatially constrained hierarchical clustering algorithm. Using 6 minutes resting state functional activity, both raw and prewhitened, we constructed two group level functional parcellations of the brain. Both atlases resulted in more homogenous regions than the commonly used AAL atlas as well as random parcellations. Separation between regions was similar to that of random parcellations for both atlases, although the functional atlas constructed using prewhitened time series had more separable regions than the random parcellations. Reproducibility of the atlases across different subject groups was moderate and did not depend on parcellation scale. Duration of time series only slightly increased the reproducibility. Reproducibility of individual level atlases were lower than group level.

Overall, spatially constrained hierarchical clustering algorithm seems to be promising for construction of the functional atlases. However, datasets with longer scan duration and more subjects are required to produce highly reproducible parcellations.

# Acknowledgements

This work was supported by NIH grants MH060662 and DA038894 and NSF grant CMMI: 1634445.